\documentclass[11pt,natbib]{article}        
\usepackage{graphicx}
\usepackage[T1]{fontenc}
\usepackage{authblk}

\usepackage[utf8]{inputenc}

\begin{document}

\title{Synaptic Noise Facilitates the Emergence of Self-Organized Criticality in the \textit{Caenorhabditis elegans}
Neuronal Network 
}

\author{Koray Çiftçi } 

  \affil{       Biomedical Engineering Department, \\
   Namık Kemal University, Tekirdag, Turkey \\
            kciftci@nku.edu.tr           
}


\date{\textit{March, 2016}}
\maketitle

\begin{abstract}
Avalanches with power-law distributed size parameters have been observed in neuronal networks. This observation might be a manifestation
of the self-organized criticality (SOC). Yet, the physiological mechanicsm of this behavior is currently unknown. Describing synaptic noise
as transmission failures mainly originating from the probabilistic nature of neurotransmitter release, this study investigates the potential
of this noise as a mechanism for driving the functional architecture of the neuronal networks towards SOC. To this end, a
simple finite state neuron model, with activity dependent and synapse specific failure probabilities, 
was built based on the known anatomical connectivity data of the nematode \textit{Ceanorhabditis elegans}. Beginning from random 
values, it was observed that synaptic noise levels picked out a set of synapses and
consequently an active subnetwork which generates power-law distributed neuronal avalanches. The findings of this study brings up the 
possibility that synaptic failures might be a component of physiological processes underlying SOC in neuronal networks.
\end{abstract}

\section{Introduction}
\label{intro}
Synaptic transmission in neurons exhibits a fair amount of randomness. This random behavior mainly originates from the probabilistic
nature of quantal release, the random nature of diffusion and chemical reactions within the synaptic cleft, and the unpredictable responses
of ligand-gated ion channels (\cite{white2000channel}). In most cases, a synapse is more likely to fail to release transmitter
in response to an incoming signal (\cite{laughlin2003communication}). The influence of noise on communication systems is rather
complex and may lead to some unexpected improvements in system capabilities (\cite{jung1991amplification}). In the same manner, the synaptic noise was shown to
advance learning capabilities of the neural network (\cite{buhmann1987influence}), maximize information
storage capacity (\cite{varshney2006optimal}),
 and improve information transmission between neural populations (\cite{gatys2015synaptic}).
 Based on this regulatory effects of synaptic noise
 in neural systems, the research described here set out to explore the influence of noise on the self organized critical behavior of
 neural systems. \par SOC has been hypothesized to be a fundamental property of neural systems (\cite{hesse2014self}).
Activity in the brain displays many different scales of organization, yet without a central executive. SOC theory, (\cite{bak1987self}), underlines the
propensity of some systems, generally consisting of large number of interacting entities, to drive themselves to criticality where they function at the edge of phase transitions. This critical regime
equips the systems with the potential to develop extended correlations in time and space, which, in the sequel, drives the
emergence of global behavior from local interactions. The existence and emergence of SOC in the brain has been
investigated both experimentally (e.g. \cite{beggs2003neuronal}, \cite{linkenkaer2001long}) and theoretically (e.g. \cite{wang2011sustained},
\cite{lin2005self}). Activity dependent synaptic plasticity has been investigated as a possible mechanism of self tuning towards SOC
(\cite{levina2008mathematical}, \cite{meisel2009adaptive}, \cite{droste2012analytical}). Neuron level synaptic plasticity generates a
network level dynamic topology (and vice versa) that provides the local neurons with global information which is critical for SOC
behavior. \par
Avalanches whose size parameters are distributed according to power-law is the main manifestation of the SOC. Power law is interesting 
because, from a qualitative perspective, although the majority of the avalanches are small in size, there is a finite possibility of 
observing middle and big sized, even reaching to the system size, avalanches. This tailors a complex interaction among network members.
In subcritical systems the interactions are mainly local whereas in supracritical systems local activations quickly spread out to the
whole system. On the other hand, in the critical systems there are both activations confined to a small region and global cascades. This
type of behavior suits very well with the observed segregation/integration balance (\cite{tononi1994measure}) and small-world regime 
(\cite{achard2006resilient}) of the neuronal networks. \par
The present study aims at exploring the potential of synaptic noise to drive neural networks towards SOC. The starting hypothesis was that
the plasticity induced by an adaptive synaptic noise process might bring out a functional network topology which exhibited neuronal avalanches.
Using a simple discrete model based on the neuronal anatomical connectivity
of \textit{Caenorhabditis elegans}, the synaptic failures were shown to be indeed essential for sustaining
the network activity at a critical regime.
 
\section{Methods}
\label{sec:1}
\subsection{The anatomical network}
\label{sec:2}
A near complete description of the nematode \textit{C. elegans} nervous system has been achieved using electron
microscopy reconstructions (\cite{varshney2011structural}), and is freely available on-line\footnote{e.g. from www.openconnectomeproject.org, www.wormatlas.org}.
\textit{C. elegans} possesses 302 neurons of which 282 are somatic and 20 are pharyngeal. Three of the somatic
neurons do not make any synapses. The remaining 279 somatic neurons make 514 gap junction connections and  2194 chemical synapses. In the 
current study, the full network formed by bringing together both types of synapses, was analyzed. Since the directionality of
gap junctions was not available these contacts were treated as bidirectional whereas the directionality
of chemical synapses were conserved. In total, this procedure generated a network of 279 neurons with 2990 directed edges. Please note that
this is the network denoted as the full network in \cite{varshney2011structural}.

\subsection{The model}

The spreading of forest fires was one of the first applications of SOC analysis (\cite{drossel1992self}). The
forest fire model has been particularly useful because it easily lends itself to describe dynamically similar albeit different systems.
Accordingly, similar models were used to describe activation spreading in a network of neurons (\cite{muller2008organization},
\cite{droste2012analytical}). The model is simplistic in the sense that a neuron, at any time, can be in any one of the 3 states:
Susceptible (S), excited (E), and refractory (R). In this study, synaptic failures were included in the model with their corresponding probability.
The evolution of the model is described by the following rules:

\begin{itemize}
 \item  A susceptible neuron can go into the excited state spontaneously with probability \textit{f}.
  \item A neuron can also be activated by the subsequent activation of one of its incoming neighbors.
  \item A synapse fails to transmit the activity with an adaptive probability \textit{g}.
 \item  After the excited state, a neuron enters into the refractory state.
 \item  The neuron can recover from the refractory state and become a susceptible neuron with probability \textit{q}.
\end{itemize}

SOC is generally inspected through observing avalanche dynamics. After a slow and long driving process, a fast avalanche event 
(in our case successive excitation of neurons) with
short duration occurs. Several orders of magnitude difference between time scales of the accumulation and
avalanche periods is a characteristic feature of SOC. This difference is reflected in the separation of scales and is usually achieved
by setting $ q \gg f$. Introducing the parameter, $\theta = q/f$,
this ratio was set to 10, 20, 50, 100, 200, 300, 500 and 1000 in this study.
In the implementation, this
corresponds to making $ \theta$ random attempts to carry refractory neurons to susceptible state, i.e. the driving phase,
which is followed by a random selection 
of a neuron for excitation. If the selected node is a susceptible node, then an avalanche starts, the constant driving stops and
the avalanche travels according to neighborhood relations. This continues until all network activations
come to an end. Accordingly, $\theta$ determines the expected time length between avalanches. The extent of the avalanche was determined
by a breadth-first search algorithm (\cite{grassberger2002critical}). \par
\begin{figure}
  \includegraphics[scale = 0.40]{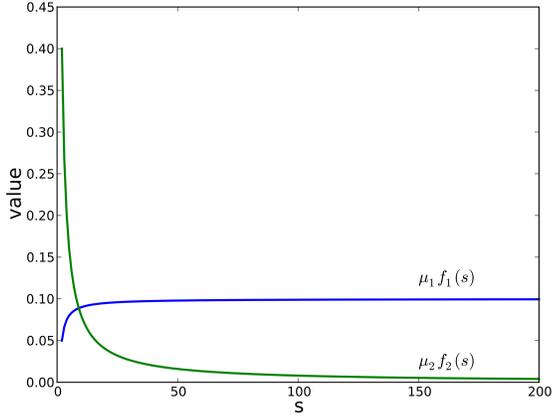}
\caption{The change in total update parameter, $\mu_k f_k$ with avalanche size (s)}
\label{fig:updateParameters}       
\end{figure}

The synaptic noise, reflected by the synaptic failure probability, \textit{g}, is the main driving force of the functional
network topology. This probability was allowed to vary depending on the avalanche formation. During an avalanche an activated node may not
be able to trigger an activation in any of its outgoing susceptible neighbors because of two reasons: That node may have been already activated
by another neighboring node or synaptic failure may not allow the transmission of activity. Accordingly, when two neighboring nodes
are both activated but the synapse between them is not the carrier of this activity, the synaptic failure probability of this synapse
is increased by,
\begin{equation}
\Delta g = \mu_1 f_1(s) (1 - g), 
\end{equation}

where, $\mu_1$ is the constant step parameter, and $f_1$ is a function depending on the avalanche size ($s$) and defined as
$\displaystyle f_1 = 1 - 1/s$.
Consequently, when two neighboring susceptible nodes are both activated via
their shared synapse, the synaptic failure probability is updates as,
\begin{equation}
 \Delta g =  - \mu_2 f_2(s) g,
\end{equation}

where, $\mu_2$ is again the constant step parameter, and $f_2$ is an avalanche size dependent function defined as
$\displaystyle f_2 = 1/s$. The determination of $f_1$ and $f_2$ is mainly heuristic: Consider that an avalanche is formed by
activated neurons and the synapses among them. Some synapses are the members of the avalanche because they propagate the activation whereas
some others are not, because they aren't able to transmit the activity because of the aforementioned reasons. We conjectured that 
the failure probability increments of an omitted synapse should grow with increasing avalanche size, whereas failure
probability decrements should get smaller with increasing avalanche sizes. In other words, small avalanches should be more selective
for successful synapses while large avalanches should be more selective for failing synapses. However, with these choices of $f_1$ and
$f_2$, the failure probability increments would always be higher than those of decrements for
avalanches occupying more than 2 neurons.
The constants $\mu_1$ and $\mu_2$ cope with this imbalance. $\mu_2$ values bigger than $\mu_1$ keep the update steps of decrements
larger than increments for a longer period. In this study $\mu_1$ and $\mu_2$ were set to $0.1$ and $0.8$, respectively. Rather than the
absolute values, the ratio is important for the performance. With these parameter settings, the change in the update parameters with
increasing avalanche size is shown in Figure ~\ref{fig:updateParameters}. It should also be noted that our selection of the
update functions enable a soft bound on the synaptic failure
probability between \textit{0} and \textit{1} and the update steps
depend on the current value.

\subsection{Simulations and data analysis}

Simulations and all related analyses were performed using Python with Numpy, Scipy (\cite{citeScipy}), and
Matplotlib, (\cite{Hunter:2007}), packages. Graph-theoretical analysis was used to relate the simulation results to the
network structure. For this purpose custom-written codes
with Python using Networkx package (\cite{hagberg-2008-exploring}) were used. Each neuron was defined as a node of the graph and each synapse was
assigned to an edge. The terms neuron/node and edge/synapse are used interchangeably throughout this paper. \par
The assessment of SOC was done mainly via fitting a power-law distribution, ($ p(s) \propto s^{-\alpha} $ ), to the two parameters
estimated from the avalanche: First, the avalanche size 
measured as the total number of neurons activated during an avalanche and second the eccentricity, i.e. the longest path length between any two
nodes of the subnetwork.  Additionally, it was checked whether the avalanche size was comparable to the 
network size. For fitting power-law distribution,
the procedure described in the seminal paper of
\cite{clauset2009power} was adopted. In summary, the scaling parameter, $\alpha$ was estimated with the method of maximum likelihood. A
Kolmogorov-Smirnov (KS) statistic was computed for this fit. After generating synthetic data sets using the same scaling parameter,
KS statistic was determined for each dataset. The null hypothesis was that our original data came from a power-law distributed variable.
To be able to reject the null hypothesis, the original KS statistic of the empirical data should be significantly higher than those of the
synthetic data. This is simply evaluated by determining what fraction
of the time the synthetic statistic is larger than the value for the empirical data. Denoting this fraction as the p-value, the null
hypothesis was rejected if $p \le 0.1$. \par

\section{Results}

\subsection{Case 1: No synaptic noise}
\begin{figure}
  \includegraphics[scale = 0.40]{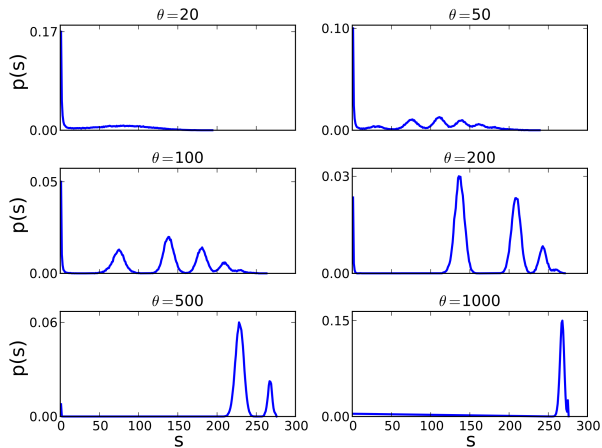}
\caption{The avalanche sizes for no synaptic noise.}
\label{fig:nonoise}       
\end{figure}

Before beginning our exposition about the effect of noise on the SOC behavior of \textit{C. elegans} network, it would be informative
to inspect the no-noise case. For this purpose the failure probability was set to \textit{0} for all synapses and the simulation runs were
repeated for 20 times. Figure ~\ref{fig:nonoise} shows the avalanche sizes for different $\theta$ values. It may be clearly observed that especially
beginning with $\theta = 50$ characteristic scale(s) for avalanches occur. Although, the relation of these avalanche scales to network
topology is a matter of interest, since the primary concern of this paper is the emergence of SOC behavior,
we will leave this topic for further studies and suffice by noting that when there is no synaptic noise the network operates in the 
supercritical regime.\par
The parameter $\theta$ determines the number of nodes that will we in the susceptible state after the driving period between the
avalanches. If we denote total number of nodes by $N$, a refractory node will be in the excited state with probability,
\begin{equation}
p = 1 - (\frac{N-1}{N})^{\theta}. 
\end{equation}

\begin{figure}
  \includegraphics[scale = 0.4]{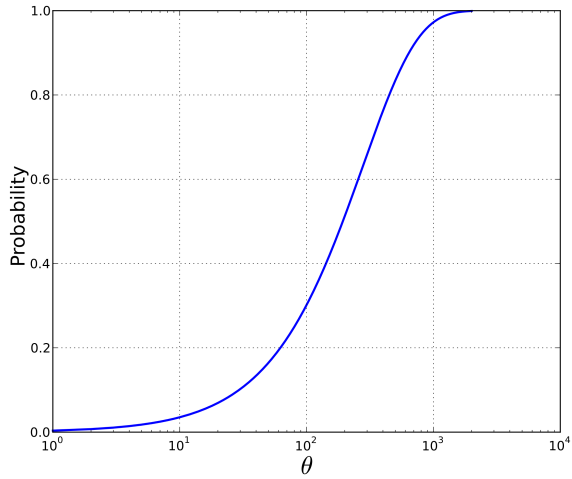}
\caption{Assuming all nodes in the refractory state, the number of susceptible nodes after $\theta$ random refractory-susceptible 
transition
attempts.}
\label{fig:theta}       
\end{figure}

Since the node selection is independent and uniformly distributed, this probability
will also give the fraction of the nodes in the refractory state (assuming all of the nodes are refractory in the beginning).
Figure ~\ref{fig:theta} quantifies this role of $\theta$ and it makes clear why at $\theta = 1000$ avalanche sizes are almost equal
to the network size: Because almost all neurons are in the susceptible state. In the actual simulations number of susceptible
neurons deviate from the numbers shown in this figure, because of the remaining susceptible neurons from the previous avalanche.
The values of Figure ~\ref{fig:theta} actually constitute lower bounds.

\subsection{Case 2: Adaptive synaptic noise}

\begin{figure}
  \includegraphics[scale = 0.4]{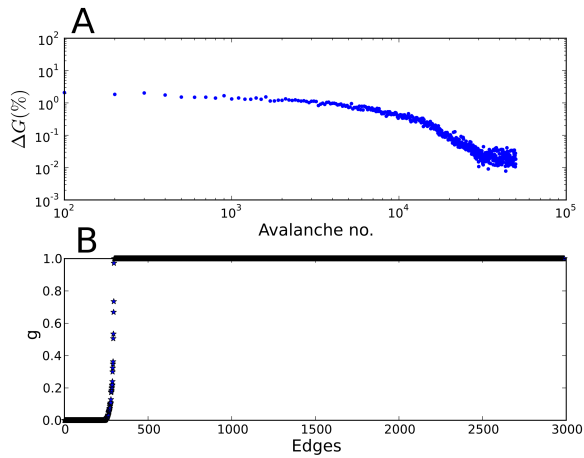}
\caption{For a single run with $\theta = 300$, A. relative total change in the synaptic failure probability values, B. (sorted) final failure probability
values.}
\label{fig:convergence}       
\end{figure}

\begin{figure*}
  \includegraphics[width=0.95\textwidth]{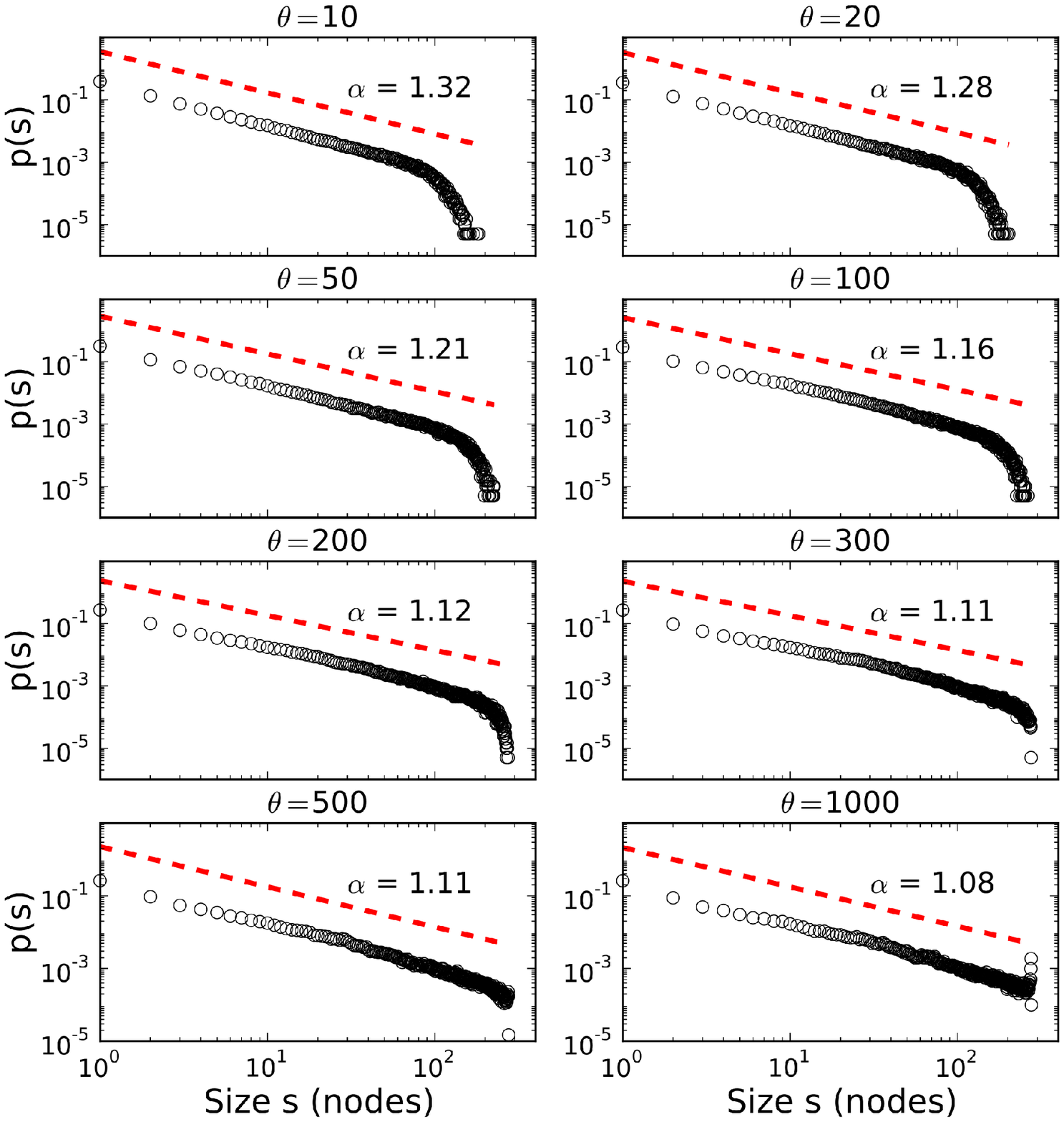}
\caption{Empirical avalanche size (\textit{s}) versus probability (\textit{p(s)}) and the power law fit, ($p(s) = s^{-\alpha}$),
graphs for different $\theta$ values.}
\label{fig:avalanches}       
\end{figure*}

\begin{figure}
  \includegraphics[scale = 0.4]{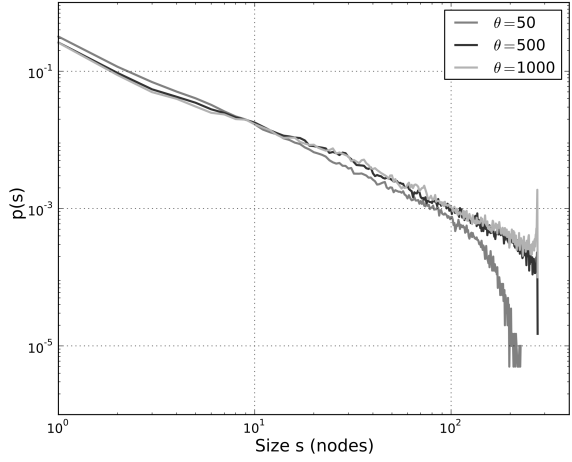}
\caption{Redrawing of the avalanche size distributions for $\theta$ equal to $50$, $500$, and $1000$.}
\label{fig:supersub}      
\end{figure}

Simulations began with all nodes in the refractory state and failure probabilities were initially assigned to
random values
from a Gaussian distribution with mean \textit{0.5} and standard deviation \textit{0.05}.
After the initialization, failure probabilities were updated for \textit{40.000} avalanches and the
convergence of these probabilities were observed.
Afterwards, statistics of \textit{10.000} avalanches were collected with fixed failure probabilities. The convergence of the failure
probabilities and the resulting values for a single run is presented in Figure \ref{fig:convergence}. Defining $G$ as the vector of
individual synaptic failure probabilities, $g$, after every 100 avalanches, the relative sum squared change in the $G$ were calculated. It may
be observed that after about \textit{30.000} avalanches convergence is attained. This convergence performance was valid for all $\theta$ values.
The figure also exhibits the final failure probabilities for the corresponding run. Most of the resulting values converge to almost 
$1$, whereas most of the remaining values converge to $0$ with few values in between. In all the simulations, less than 400 (out of 2990)
probability
values converge to values less than $1$. Although values close to $1$ actually pruned away the corresponding edges,
no node was excluded from the resulting network.
Figure \ref{fig:avalanches} demonstrates
the avalanche sizes measured as the total number of activated neurons, for different $\theta$ values. For small $\theta$ values,
($<100$), the avalanche sizes begin to diverge earlier from the power law which is a manifestation of the subcritical dynamics.
For $\theta$ values over $100$ and critical regime is attained (\textit{KS} statistics with $p = 0.1$). 
The cut-off observed in the avalanche size is due to the finite size of the network. For values close to $1000$ the avalanche distribution
begins to exhibit a sharp positive deflection close to the network size. We do not conceive this as an indication of the network entering
into the supracritical regime, but rather again as a result of the limiting effect of the network size. To make this point clear, we
carried the $\theta$ value to its utmost level, so as to make all neurons susceptible after each avalanche. The result was qualitatively
similar and this observation corroborated our conjecture on the limiting effect of the network size. 
The difference between these three behaviors is more evidenced in Figure \ref{fig:supersub}. \par
\begin{figure}
  \includegraphics[scale = 0.4]{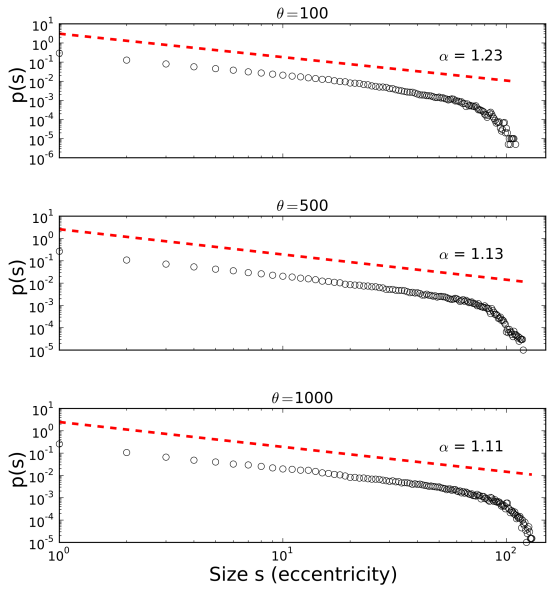}
\caption{The eccentricity (maximum path length) for the avalanche subgraph for different $\theta$ values. }
\label{fig:eccentricity}       
\end{figure}

The second size parameter investigated for the power-law was eccentricity. The activated nodes during an avalanche and the active
synapses among them form a subgraph of the original graph. This subgraph was extracted at each avalanche event and the eccentricity, 
maximum path length
between any two nodes, in this reduced network was determined. The results are presented in Figure \ref{fig:eccentricity}. The results
are again indicative of a critical behavior in the \textit{C. elegans} network. Noting that the full network has an eccentricity
value of \textit{7}, observed big eccentricity values point out to the long chains of neurons shaped by the synaptic noise levels.

\section{Discussion}

Two assertions formed the basis of this study: First, noise controls the level of activity and hence is of functional importance
for the nervous system \cite{buhmann1987influence}, and second, neuronal avalanches produced by SOC might be a
new mode of network activity (\cite{beggs2003neuronal}).
Motivated by these hypotheses this study set out to explore the interactions between these
two processes for the \textit{C. elegans} neuronal network. The main finding is that an adaptive synaptic noise lends
itself as a possible mechanism that drives the network towards SOC. Considering that there are various criticisms towards SOC theory, (e.g. \cite{frigg2003self}).
the results presented here do not supply any positive or negative evidence towards the existence SOC, rather it claims that
if SOC is indeed a mode of network activity, synaptic noise might be a mechanism that allows the anatomical network to generate
dynamical functional topologies for SOC behavior. \par

Experimental studies showed that synaptic failure probability is generally above 0.5 and can be well in excess of 0.9
(\cite{allen1994evaluation}, \cite{hessler1993probability}). The noise arising from the probabilistic nature of neurotransmitter release is actually an
important mechanism of plasticity (\cite{rosenmund1993nonuniform}). The plasticity in terms of spike timing, its update rules, parameters
have been explored both experimentally and
analytically (\cite{van2000stable}). Consequently, we have a large set of feasible rules and parameters which allow the researchers
to search for neurobiologically realistic determinants of SOC based on spike time dependent plasticity (\cite{rubinov2011neurobiologically}).
Synaptic noise was also shown to have a nonrandom component which modulates neuron function (\cite{faure1997nonrandom}). In this study,
an update rule for this noise is proposed and tested on the neuronal network of \textit{C. elegans}. The inclusion of the
avalanche size as a synaptic failure update parameter constitutes 
the weakest point in all our modeling effort. The question of how a synapse gets information about the avalanche size is left 
unanswered in this study. A neuron gets information from other neurons that it has direct contact. Hence, a link between avalanche
size and local activity profile should be sought. Providing global information to neural elements has also been a problem for neural models
assigning spreading cascades along shortest paths. In a recent study hubs and central pathways were shown to be dominating this shortest path activation
(\cite{mivsic2015cooperative}).\par
It should also be reiterated that the model used in this study is over simplistic.
Nevertheless, it has been already been demonstrated that this type of simple models produced results in excellent
agreement with more realistic simulations (\cite{messe2015closer}). By abstracting away microscopic details
this type of simple models emphasize the emergence of global patterns from local neural interactions (\cite{mivsic2015cooperative}). Moreover,
similar models were efficiently employed to model the activity propagation in neural networks (\cite{stam2015relation}).
The inclusion of the noise term actually drives our model closer to a variant of the forest fire model, in which immunity against fire 
is given to trees with some fixed probability (\cite{clar1996forest}). \par 

Avalanches were hypothesized to be transient formation of cell assemblies (\cite{plenz2007organizing}). They represent spatially irregular
patterns of propagated synchrony which are stable and exhibit recurrence. The neuron chains shaped by the synaptic noise can be considered
as a manifestation of these assemblies. As a post-hoc investigation, the distribution of failure probabilities were analyzed to understand
whether this distribution was similar across simulations. There were no significant correlations among the distributions. This observation
indicates that neural networks might include many different overlapping functional assemblies capable of generating complex activation patterns.\par

Robust statistical assessment of power-law statistics is problematic with finite size systems (\cite{taylor2013identification}). In our model
setting, since the driving of the neurons, i.e. transition from refractory to susceptible, stops during an avalanche, it is not 
possible for a neuron to reactivate within the same avalanche. This means that the maximum avalanche size (in terms of the number of activated
neurons) is strictly limited by the network size. It is known that finite size systems exhibit a cut-off dictated by the system size
(\cite{plenz2007organizing}). This was also evident in our results. But, for the critical regime a strong power-law behavior was observed
up to almost the network size. \par

Same model without the noise component and avalanche type activation revealed that activity in the neural network of \textit{C. elegans}
were dominated by central hub nodes (\cite{muller2008organization}). The well defined activation sizes in our no-noise networks should
be reconsidered in the future within this perspective. This behavior might also be conceived as a manifestation of network-shaped self 
organization (\cite{hutt2014perspective}). It was also observed that \textit{C. elegans} neural network operated at a critical region
rather than a certain critical point (\cite{moretti2013griffiths}). In the present work, the existence of many different synaptic noise distributions
each giving rise to critical behavior might also be evaluated in the same light.


\bibliographystyle{abbrv}      
\bibliography{synapticNoise_SOC2}   

\end{document}